\documentclass[showpacs,twocolumn,amsmath,superscriptaddress,pre]{revtex4}
\usepackage{epsfig}  

\begin{document}

\preprint{EV/126-05-2006}

\title{The magnetization-driven random field Ising model at $T=0$ }

\author{Xavier Illa}
\email{xit@ecm.ub.es}
\affiliation{ Departament d'Estructura i Constituents de la Mat\`eria,
  Universitat de Barcelona \\ Diagonal 647, Facultat de F\'{\i}sica,
  08028 Barcelona, Catalonia}
\author{Martin-Luc Rosinberg}
%
%
\affiliation{Laboratoire de Physique Th\'eorique de la Mati\`ere
  Condens\'ee, Universit\'e Pierre et Marie Curie, 4 Place Jussieu,
  75252 Paris, France}
\author{Prabodh Shukla}
%
%
\affiliation{ Physics Department, North Eastern Hill University,
  Shillong 793003, India}
\author{Eduard Vives}
%
%
\affiliation{ Departament d'Estructura i Constituents de la Mat\`eria,
  Universitat de Barcelona \\ Diagonal 647, Facultat de F\'{\i}sica,
  08028 Barcelona, Catalonia}

\date{\today}

\begin{abstract}
  We study the hysteretic evolution of the random field Ising model
  (RFIM) at $T=0$ when the magnetization $M$ is controlled externally
  and the magnetic field $H$ becomes the output variable. The dynamics
  is a simple modification of the single-spin-flip dynamics used in
  the $H$-driven situation and consists in flipping successively the
  spins with the largest local field. This allows to perform a
  detailed comparison between the microscopic trajectories followed by
  the system with the two protocols. Simulations are performed on
  random graphs with connectivity $z=4$ (Bethe lattice) and on the
  3-D cubic lattice.  The same internal energy $U(M)$ is found with
  the two protocols when there is no macroscopic avalanche and it does
  not depend on whether the microscopic states are stable or not. On
  the Bethe lattice, the energy inside the macroscopic avalanche also
  coincides with the one that is computed analytically with the
  $H$-driven algorithm along the unstable branch of the hysteresis
  loop.  The output field, defined here as $\Delta U/\Delta M$,
  exhibits very large fluctuations with the magnetization and is not
  self-averaging. Relation to the experimental situation is discussed.

\end{abstract}


\maketitle

\section{Introduction}
 
The random-field Ising model (RFIM) is one of the simplest model to
study the combined effects of interaction and disorder in many-body
systems.  In particular, the response of the RFIM to a slowly varying
magnetic field at zero temperature\cite{SDP2004} illustrates the
athermal dynamical behavior observed in several experimental systems
in condensed matter physics such as disordered ferromagnets,
superconductors, martensitic materials, etc. This response is
characterized by avalanches and rate-independent hysteresis. Recently,
the model has also been transposed to the context of finance and human
behavior\cite{MB2005}.

The aim of the present work is to study the $T=0$\ RFIM in a situation
that has not been considered so far, when one varies the overall
magnetization and not the magnetic field (which then becomes a derived
quantity that we will call the ``output'' field).  More generally, we want
to describe the behavior of athermal systems under control of the {\it
  extensive} variable conjugated to the intensive force. This concerns
for instance the stress-strain curves in shape-memory materials that
are usually obtained by controlling the deformation of the sample and
measuring the induced stress\cite{OW1998}. One also uses a feedback
control that imposes a constant variation of the magnetic flux in the
case of ferromagnets with a very steep magnetization
curve\cite{G1977}.

For a system at equilibrium, it is of course equivalent to control the
force or the conjugated variable: the system follows a well-defined
curve which corresponds to the minimum of the energy or the
free-energy. This curve may be continuous or discontinuous, as is the
case at a first-order phase transition. The situation is more
complicated when thermal fluctuations are too small to overcome the
energy barriers and the system remains far from thermodynamic
equilibrium on the experimental time scale. It then follows a
metastable, history-dependent path, and there is no reason for
observing the same behavior with the two protocols. In fact, there is
experimental evidence that hysteresis loops obtained by varying
extensive variables display bending-back trajectories (with a
so-called yield point), and large fluctuations in the measured force
(or field)\cite{G1977,BMPRV2006}.

In order to simulate this situation with the $T=0$ RFIM, one needs to
introduce a dynamical rule that states how to flip the spins as the
magnetization is changed. There are of course different ways of
locally minimizing the energy and the choice for the dynamics is not
unique, even if one imposes a deterministic rule so to get the same
result when repeating the simulation. In this work, we propose to
modify the standard single-spin-flip dynamics in a {\it minimal} way,
so that the new dynamics may be considered as the
``magnetization-driven'' version of the dynamics used in the
field-driven case\cite{SDKKRS1993}. The main advantage is that there
is a close connection between the microscopic trajectories followed by
the system with the two protocols and the results for the macroscopic
quantities (for instance the internal energy) can be readily compared,
 
Another and more delicate issue concerns the definition of the
magnetic field as an output variable. The solution that we adopt is
again very simple but cannot be considered as fully satisfactory.  In
another recent work\cite{IRV2006}, a different approach was proposed,
extending the study to finite temperatures so to define the field as a
Lagrange multiplier. Comparison between these two approaches is
discussed below. Part of our study is performed on a Bethe lattice
with connectivity $z=4$ (or, equivalently, on random graphs with the
same connectivity). This is to benefit from the fact that an almost
complete analytical description is available in the field-driven
case\cite{DSS1997,SSD2000,IOV2005}. Comparing our simulation data with
these exact results will help in understanding the similarities and
differences between the two protocols.

In section \ref{Model}, we review the model in the usual field-driven
situation and introduce the modifications in the dynamics so to
describe the magnetization-driven case.  The simulation results for
the Bethe lattice are discussed in section \ref{Bethe} and those for
the 3-D cubic lattice in section \ref{3d}. We summarize our main
findings and conclude in section \ref{Discussion}.

\section{Model}
\label{Model}

The RFIM with single-spin-flip local relaxation dynamics was
specifically introduced for studying the $H$-driven situation. It is
thus usually formulated from a microscopic Hamiltonian $\cal H$ that
corresponds to the magnetic enthalpy. For the present study, it is
convenient to first introduce the internal energy $\cal U$:
\begin{eqnarray}
{\cal U} & = & -\sum_{\langle i,j \rangle} S_i S_j -
\sum_i h_i S_i  \label{U}
\end{eqnarray} 
where $S_i=\pm 1$ are spin variables defined on the sites $i=1,\dots,
N$ of a lattice and the first sum extends over all nearest-neigbor
pairs (the coupling constant is taken as the energy unit and set to
unity). The random fields $h_i$ are i.i.d. variables sampled from the
Gaussian distribution $\rho(h) =\exp(-h^2/2
\sigma^2)/\sqrt{2\pi}\sigma$ with standard deviation $\sigma$. The
enthalpy $\cal H$ is then defined as
\begin{equation}
{\cal H}= {\cal U} -HM
\end{equation}
where $M=\sum_i S_i$ is the overall magnetization. In the following,
we consider two types of lattice: a 3-D cubic lattice and a Bethe
lattice with connectivity $z=4$.  In the first case, numerical
simulations are performed on finite lattices of size $N=L\times
L\times L$ with periodic boundary conditions. In the second case, they
are performed on random graphs with fixed connectivity $z=4$ which
provide a convenient realization of the Bethe lattice in the
thermodynamic limit.

\subsection{$H$-driven dynamics}

The standard $H$-driven dynamics consists in locally minimizing the
enthalpy $\cal H$. As the external field $H$ is changed, each spin is
aligned with its total local field $f_i+H$, where
\begin{equation}
f_i = \sum_{j/i} S_j +h_i
\end{equation}
and the summation is over all the $z$ neighbors $j$ of $i$. A
configuration $\{S_i\}$ is then (meta)stable when all the spins
satisfy the condition.
\begin{equation}
S_i =\operatorname{sign}(f_i+H)
\label{stabH}
\end{equation}
One usually starts the metastable evolution with $H=-\infty$ and all
spins $S_i=-1$. $H$ is then increased until the total local field
vanishes at a certain site. This first occurs for the spin with the
largest random field, $h_i^{max}$.  This spin is then flipped, which
in turn changes the local field at the neigbors and may trigger an
avalanche of other spin flips.  The avalanche stops when a new stable
configuration is reached. $H$ is then increased again until a new spin
becomes unstable and the evolution continues until all the spins flip
up. The upper half of the hysteresis loop is obtained in a similar way
by decreasing the field from $+\infty$ to $-\infty$. Note that the
external field $H$ is kept constant during an avalanche, which
corresponds to a complete separation of time scales between the
driving mechanism and the internal relaxation of the system (the
dynamics is then referred to as ``adiabatic''). Because the
interactions are purely ferromagnetic, the dynamics has also some
remarkable properties: it is abelian\cite{SSD2000} (the order in which
unstable spins are flipped during an avalanche is irrelevant for
determining the final state) and it satisfies return-point
memory\cite{SDKKRS1993}. An important feature is the existence of a
critical amount of disorder $\sigma_c$ below which the hysteresis
loops are discontinuous in the thermodynamic limit, the jump in the
magnetization corresponding to the occurence of a macroscopic
avalanche\cite{SDKKRS1993}. One has $\sigma_c\approx 2.2$ for the
cubic lattice\cite{Perkovic1999,PV2003} and $\sigma_c=1.781258...$ for
the Bethe lattice with connectivity $z=4$\cite{DSS1997}.

\begin{figure}[htb]
\begin{center}
  \epsfig{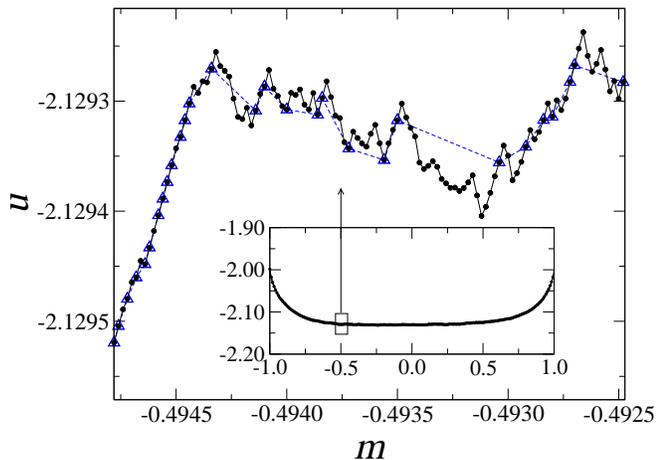}
\end{center}
\caption{\label{FIG1}(Color on line) Comparison between the $H$-driven  and $M$-driven  trajectories in the energy-magnetization plane ($u={\cal U}/N$ is the internal energy per spin). Data correspond to a single disorder realization with $\sigma=2$ on a random graph with connectivity $z=4$ ($N=10^5$). The triangles represent the states visited by the $H$-driven dynamics which are separated by avalanches (dashed lines). The dots are the states visited by the $M$-driven dynamics. }
\end{figure}

Fig.~\ref{FIG1} shows an example of an $H$-driven metastable evolution
on a random graph with connectivity $z=4$. For the sake of comparison
with the $M$-driven protocol that is introduced in the next section,
we plot the internal energy per spin $u={\cal U}/N$ as a function of
the magnetization per spin $m=M/N$ (both quantities being parametrized
by the external field $H$).  The metastable states
$\{S_i\}_1,\{S_i\}_2,\{S_i\}_3$,... visited by the dynamics are
represented by triangles while the dashed lines in-between indicate
the avalanches. Note that the total number of states when $H$ is
varied from $-\infty$ to $+\infty$ depends on the disorder strength
and on the particular realization of the random fields.  Typically,
there are only a few states when $\sigma$ is small (most avalanches
are large) whereas the number of states approaches its upper limit $N$
when $\sigma$ is large.

Finally, we want to stress that the energetic barriers between the
metastable states are strictly defined by the dynamics. In fact, the
very definition of the metastable states (i.e. the stability rule
(\ref{stabH})) cannot be separated from the use of the
single-spin-flip dynamics. It has been shown recently that a slightly
better minimization of the enthalpy (obtained by allowing also
simultaneous flips of nearest-neigbor spins) yields much thinner
hysteresis loops while not changing the critical behavior of the
system\cite{Vives2005}.

\subsection{$M$-driven dynamics}

We now define an irreversible dynamics for the case where the
magnetization of the system is changed externally. There is no
external field and the potential that has to be minimized (at least
partially) is the internal energy $\cal U$.  Our goal is to generate a
sequence of states $\{S_i\}_1,\{S_i\}_2,\{S_i\}_3$,... when $M$ is
increased from $M=-N$ to $M=+N$ by elementary steps $\Delta M=2$.  As
noted in the introduction, we want this dynamics to be as close as
possible to the single-spin-flip dynamics used in the $H$-driven case.
For instance, we require that the two driving mechanisms become
equivalent when the spins behave independently and the hysteresis
vanishes (either because the coupling constant is zero or $\sigma
\rightarrow \infty$). In this limit, one must thus flip, for each
value of $M$, the spin with the largest random field, $h_i^{max}$. In
the general case, we propose to use the simplest ``extremal''
dynamics: the spins are flipped one by one (like in the $H$-driven
case) and, for each value of $M$, one chooses the spin that {\it most}
decreases or, at least, {\it less} increases the internal energy. This
is the spin with the largest local field, $f_i^{max}$, and the
corresponding change in the energy is $\Delta {\cal U}=-2 f_i^{max}$.
After the spin has been flipped, the local fields $f_i$ at the
neigbors are updated and the same rule is applied until all spins are
flipped. One obtains a different sequence of states when starting from
$M=+N$ and decreasing the magnetization, which yields an hysteresis
loop. It may be remarked that this new dynamics bears some similarity
with the ``extremal'' dynamics used in simple models of self-organized
criticality (see e.g. Ref.\cite{S1992}). However, in the present case,
one never reaches a statistically stationary state because each spin
in the system flips only once and $m$ evolves between $-1$ and $+1$.

By construction, the total number of states visited by the dynamics is
now $N$ and the crucial feature is that this sequence of states
contains {\it all} the $H$-driven metastable states as a subsequence.
This is due to the abelian property of the $H$-driven dynamics and can
be easily understood by noticing that i) the two dynamics start with
the same initial state (with all $S_i=-1$ or $+1$), and ii) the spins
that are flipped successively within the $M$-driven dynamics are
either those which trigger an $H$-driven avalanche or those which are
involved in this avalanche. In other words, the dynamical rule that
has been chosen generates a sequence of states that are obtained by
flipping {\it in a certain order} the spins involved in the $H$-driven
avalanches. This is illustrated in Fig.~\ref{FIG1} which shows the
sequences of states obtained with the two dynamics in the $u-m$ plane.
One can see that the $M$-driven trajectory is a sort of random walk
that joins the metastable states belonging to the $H$-driven
trajectory.

A more problematic (but separate) issue concerns the definition of the
output field $H$ associated to the changes in the magnetization. As
will be discussed below in more detail, one difficulty is that many of
the states visited by the dynamics are not metastable. This means that
is not possible to find a field that allows for the condition
(\ref{stabH}) to be satisfied for all spins. This is because the local
field $f_i$ at some spins down is larger than the local field at some
spins up (it is easy to see from Eq.(\ref{stabH}) that the condition
for a microscopic configuration $\{S_i\}$ to be metastable at some
field $H$ is that $f_i^{min}$, the minimum value of the local field
among the spins up, is larger than $f_i^{max}$, the maximum value of
the local field among the spins down). In this respect, the present
situation is totally different from the one considered in
Ref.~\cite{IRV2006} where all the states obtained with the $M$-driven
dynamics are stable. An additional difficulty is that there is no
obvious way to define an intensive quantity conjugated to $M$, playing
the same role as the Lagrange parameter introduced in Ref.~\cite{IRV2006}.
The simple solution that we propose is to define the field in such a
way that the work needed to go from the state at $M$ to the state at
$M+\Delta M$ is minimal. The field thus identifies with the internal
force,
\begin{equation}
H(m)\equiv \Delta {\cal U}/\Delta M=-f_i^{max}(m) \ .
\label{defH}
\end{equation}
(In order to facilitate the comparison with the $H$-driven dynamics we
use the same notation for the external and the output field. For the
metastable states that are common to the two dynamics, the field
defined by Eq. (\ref{defH}) is exactly the external field at which
these states become marginally stable.)
\begin{figure}[htb]
\begin{center}
  \epsfig{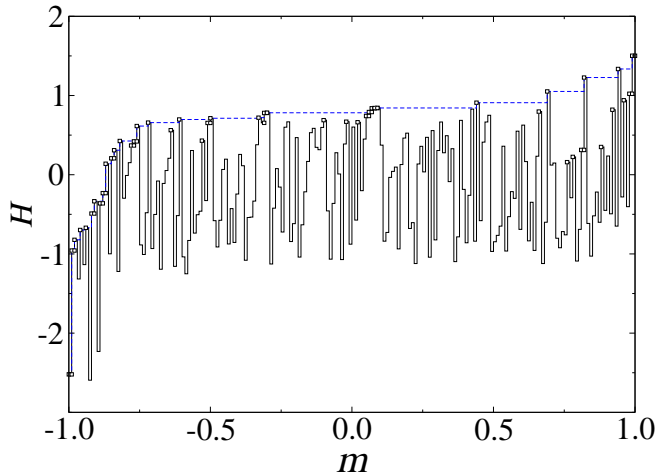}
\end{center}
\caption{\label{FIG2}(Color on line)   
  Comparison between the $H$-driven (dashed line) and $M$-driven
  (solid line) trajectories in the field-magnetization plane.  Data
  correspond to a single disorder realization with $\sigma=2$ on a
  random graph with connectivity $z=4$ ($N=200$). The symbols
  represent the metastable states according to the single-spin-flip
  dynamics (some of them do not belong to the $H$-driven trajectory,
  as discussed in section III A).  }
\end{figure}

$H(m)$ is not a monotonously increasing function of the
magnetization. In fact, as can be seen in Fig.~\ref{FIG2} in the case
of a very small system, it strongly fluctuates with $m$. This is a
quite different behavior from the one observed for the magnetization
in the $H$-driven case. Note however that one can easily deduce the
$H$-driven trajectory from the $M$-driven one (when the magnetization
is put on the horizontal axis): it is just the envelope function that
tracks the increasing maxima of $H(m)$.

A direct consequence of Eq.(\ref{defH}) is that the $M$-driven
dynamics does not yield any dissipation. Indeed, since the work
$H\Delta M$ is just equal to the variation of the internal energy, the
area of any closed loop (that goes back to the same microscopic state)
is zero. This is to be contrasted with the situation in the $H$-driven
case in which the work is larger than $\Delta {\cal U}$ inside the
avalanches.  Although the experimental hysteresis loops obtained in
$M$-driven conditions have a much smaller area than the $H$-driven
loops\cite{G1977,BMPRV2006}, it is not true that the dissipation is
zero. We shall come back to this issue in section \ref{Discussion}
where we discuss some possible modifications in the definition of the
field so to avoid this ``unphysical'' feature.

\subsection{Fluctuations and self-averaging}

The two preceding algorithms allow to simulate a finite system for a
given realization $\{h_i\}$ of the random fields. Comparison with
experiments should be performed by considering the limit $N
\rightarrow \infty$. It is then desirable that the results are
self-averaging, i.e. that they do not depend on a particular
realization of the disorder in the thermodynamic limit.

In this respect, the situation is different when one is controlling
the external field $H$ or the magnetization $M$. In the former case,
one can divide a macroscopic system into a large number of macroscopic
subsystems that are all submitted to the {\it same} external field.
Then, according to a standard argument\cite{B1959}, away from
criticality, the value of the density of any extensive quantity on the
whole system (for instance the magnetization $M(H)$) is equal to the
average of the (independent) values of this quantity over the
subsystems. According to the central limit theorem, this quantity is
distributed with a Gaussian probability distribution and (strongly)
self-averaging\cite{BH1988}. On the other hand, in the latter case,
one cannot decompose a system into subsystems having the same
magnetization and the standard argument does not apply. This implies
that i) one must carefully study the behavior of the sample-to-sample
fluctuations of an observable as the system size increases so to
conclude whether or not this observable is self-averaging, and ii) one
must be cautious in giving a physical meaning to the average over
disorder.

In the following, we analyze the self-averaging character of an
observable $X$ by performing histograms over many disorder
realizations for a given size $N$, so to estimate the probability
distribution $P_N(X)$. We then study the behavior of the variance
$V_X= \langle X^2 \rangle_M - \langle X \rangle_M^2$ as $N$ increases
(here $\langle .\rangle_M$ denotes the average over disorder at
constant $M$, which has to be distinguished from $\langle .\rangle_H$,
the average over disorder at constant $H$). $X$ is (strongly)
self-averaging if $V_X \sim 1/N$ when $N\rightarrow \infty$.

\section{Results for the $z=4$ Bethe lattice}
\label{Bethe}
In this section we present the numerical results obtained by
simulating the $M$-driven dynamics on random graphs with fixed
connectivity $z=4$. Since small loops are rare in these graphs (their
typical size is of order $\log N$), the results in the large-$N$ limit
are expected to converge to the results on a true Bethe lattice, i.e.
in the deep interior of a Cayley tree.  For the sake of completeness,
we first recall the analytical expressions for the average
magnetization per spin $\langle m \rangle _H$ and the average internal
energy per spin $\langle u \rangle_H $ as a function of the external
field $H$\cite{DSS1997,IOV2005}:

\begin{equation}
\langle m \rangle_H =1-2 \sum_{n=0}^{z} {z \choose n} 
P^{\ast n} (1-P^{\ast})^{z-n} p_n
\label{mag}
\end{equation}
\begin{eqnarray}
& & \langle u  \rangle_H =  -\frac{1}{2}z + 2 \sum_{n=0}^{z} {z \choose n}
P^{\ast n}[1-P^{\ast}]^{z-n} \nonumber \\ & & \times [ n(1-p_n) +
\sigma ^2\rho(z-2n-H)] 
\label{ener}
\end{eqnarray}
where the quantity $P^{\ast}$ is solution of the equation
\begin{equation}
P^{\ast}=\sum_{n=0}^{z-1} {z \choose n} 
P^{\ast n} (1-P^{\ast})^{z-1-n} p_n
\label{past}
\end{equation}
and the functions $p_n(H)$ ($n=0,1...,z)$ are integrals of the
Gaussian distribution,
\begin{equation}
p_n=\int_{-J(2n-z)-H}^{+\infty} \rho(h) dh \ .
\end{equation} 
(Eq. (7) is obtained by summing the different contributions to the
internal energy computed in Ref.\cite{IOV2005}.) In the following, we
shall also use the expression for the probability (per spin) that an
avalanche is initiated when the field is increased from $H$ to $H+dH$.
It is defined as $G(H)dH$ with
\begin{equation}
G(H)= \sum_{n=0}^{z} {z \choose n}
P^{\ast n}[1-P^{\ast}]^{z-n} \rho(z-2n-H)
\label{RH}
\end{equation}
(this expression corresponds to Eq. (14) in Ref.\cite{SSD2000} with
$x=1$).

\begin{figure}[htb]
\begin{center}
  \epsfig{file=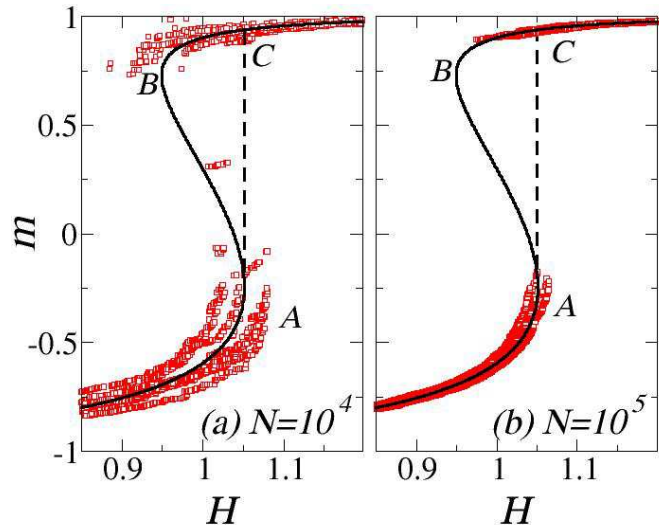,width=8.6cm,clip=}
\end{center}
\caption{\label{FIG3} (Color on line) Ascending branch of the $H$-driven hysteresis loop on 
a Bethe lattice with connectivity $z=4$ for $\sigma=1.6$. The solid line 
corresponds to the solution of Eqs. (\ref{mag}) and \ref{past}). 
The symbols represent all the metastable states obtained along the $M$-driven 
trajectories for $10$ disorder realizations on random graphs of size $N=10^4$ 
(a)  and  $10^5$ (b). }
\end{figure}

When computing $P^{\ast}(H)$, it is important to take into account the
fact that Eq. (\ref{past}) has three real roots in a certain range of
$H$ below $\sigma_c$. Then, the magnetization curve obtained from Eq.
(\ref{mag}) has an S-shape behavior\cite{DSS1997}, as illustrated in
Fig.~\ref{FIG3}. The correct physical solution that gives the lower
branch of the hysteresis loop corresponds to the smallest root. For a
certain value of $H$, this root is not real anymore, $P^{\ast}(H)$
jumps to the largest root and there is a discontinuity in the
magnetization curve associated to the occurence of an infinite
avalanche. In this context, the intermediate, unstable branch of the
S-shape curve (from points A to B in the figure) has no physical
meaning (on the other hand, the branch BC can be reached via
first-order reversal curves obtained from the descending branch of the
hysteresis loop, as noted in Refs.\cite{DRT2005,ISV2006}).

\subsection{Fraction of stable states along the $H$-driven and $M$-driven trajectories}

As can be seen in Fig.~\ref{FIG2}, which corresponds to the simulation
of a small system, there are a few states along the $M$-driven
trajectory that are metastable although they do not belong to the
$H$-driven subsequence (this could be seen as well in the $u-m$
diagram since this property does not depend on the definition of the
field). Of course, the fact that the state visited by the dynamics is
stable or not depends on the disorder realization. Is is therefore
useful to introduce the quantity $Q(m)$ that represents the average
fraction of states that are stable. As shown in Fig.~\ref{FIG3}, these
additional metastable states appear less and less frequently as the
system size increases and there is strong numerical evidence that they
completely disappear in the thermodynamic limit. Therefore, when
$N\rightarrow \infty$, $Q(m)$ also represents the average fraction of
metastable states along the $H$-driven trajectory.

\begin{figure}[htb]
\begin{center}
  \epsfig{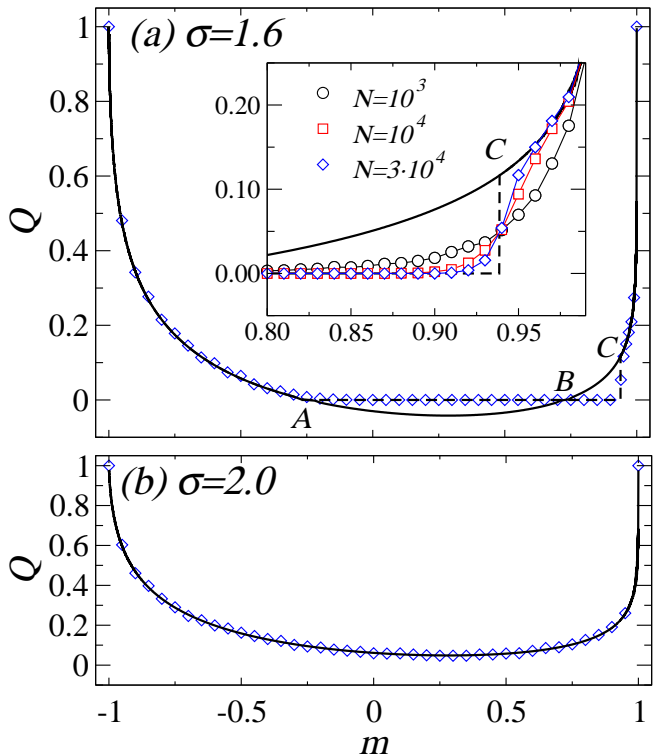}
\end{center}
\caption{\label{FIG4}(Color on line) Fraction $Q(m)$ of stable states along the ascending 
branch of the hysteresis loop for a Bethe lattice with $z=4$. The points
 A, B and C are the same as in Fig.~\ref{FIG3}. The symbols are the results 
of the numerical simulation of the $M$-driven algorithm. The solid line 
corresponds to the analytical expression (\ref{QM}) and the dashed line 
indicates the infinite avalanche below $\sigma_c$. The inset in (a) shows 
that $Q(m)$ converges to $0$ between the points B and C in the thermodynamic 
limit. Simulation data, indicated by different symbols are joined by guides 
to the eye.}
\end{figure}

The results of the simulations with the $M$-driven dynamics below and
above $\sigma_c$ and different system sizes are shown in
Fig.~\ref{FIG4}. For $\sigma>\sigma_c$, it is found that $Q(m)$ is
minimum in the range of $m$ that corresponds to the steepest part of
the $H$-driven magnetization curve where the avalanches are the
largest. For $\sigma<\sigma_c$, the interval where $Q(m)=0$ exactly
corresponds to the range of the infinite avalanche (including in the
portion BC of the magnetization curve, as shown in the inset of
Fig.~\ref{FIG4}).  The fact that $Q(m)$ is {\it strictly} smaller than
$1$ (except for $m=\pm 1$) deserves some explanation. With the
$H$-driven algorithm, there is indeed a certain probability, for a
finite system, that a given value of $M$ corresponds either to an
horizontal portion of the magnetization curve (a metastable state) or
to a vertical jump (an avalanche). Although the magnetization curve is
continuous in the thermodynamic limit (except for the jump below
$\sigma_c$), the probability of ``hitting'' a metastable states
remains smaller than $1$ when $N\rightarrow \infty$. In other words,
$Q(m)$ tracks the random presence of ``holes'' in the magnetization
curve that correspond to the avalanches. This suggests that $Q(m)$ is
related to the probability of having an avalanche between $H$ and $H+
dH$ (where $H$ is the field corresponding to $m$ in the thermodynamic
limit).  This probability is given by the quantity $G(H)$ defined by
Eq.  (\ref{RH}) and the seeked relation is
\begin{equation}
Q(m) = 2 G(H) \frac{dH}{dm} 
\label{QM}
\end{equation} 
where $dH/dm$ is the inverse slope of the magnetization curve in the
thermodynamic limit, a quantity that is easily computed from Eq.
(\ref{mag}). One can see in Fig.~\ref{FIG4} that the agreement between
the simulations and the analytical formula is indeed very good. The
proof of Eq. (\ref{QM}) relies on the assumption that the occurrence
of avalanches (as $H$ is monotonously increasing) corresponds to a
non-stationary Poisson process\cite{SDKKRS1993}. For a finite system
of size $N$, an avalanche then occurs in the interval $dH$ with a rate
$dP/dH= N G(H)$ (recall that $G(H)dH$, as calculated in
Ref.\cite{SSD2000}, is a probability per spin). The mean range of
stability $\langle \Delta H \rangle_H$ of a metastable state before an
avalanche occurs is given by the inverse of the rate, i.e.
\begin{equation}
\langle \Delta H  \rangle_H \sim \frac{1}{N G(H)} 
\label{MDH}
\end{equation}
(Note that this quantity becomes infinitesimal in the thermodynamic
limit.) Since only metastable states contribute to the variation of
$H$ in the interval $M,M+2$ (the field is kept constant during an
avalanche), the average slope of the magnetization curve in the
thermodynamic limit is given by
\begin{equation}
\frac{dH}{dm}=Q(m)\lim_{N\rightarrow\infty}N\frac{\langle \Delta H  \rangle_H}{\Delta M}
\end{equation}
which yields Eq. (\ref{QM}).

It is interesting to remark that Eq.~(\ref{QM}) gives a finite,
positive value of $Q(m)$ between points B and C in Fig.~\ref{FIG4}(a)
when one computes $m(H)$ using the largest root of Eq.~(\ref{past}).
Indeed, as already noted, this part of the $H$-driven hysteresis loop
can be reached by an appropriate field history starting from
saturation: this implies that the fraction of metastable states is not
zero. On the other hand, one gets a meaningless negative value for
$Q(m)$ between points A and B and the correct physical result
$Q(m)=0$, that states that all configurations are unstable, is
recovered by setting $dm/dH\rightarrow\infty$ in Eq.~(\ref{QM}).

\subsection{Internal energy}
\begin{figure}[htb]
\begin{center}
  \epsfig{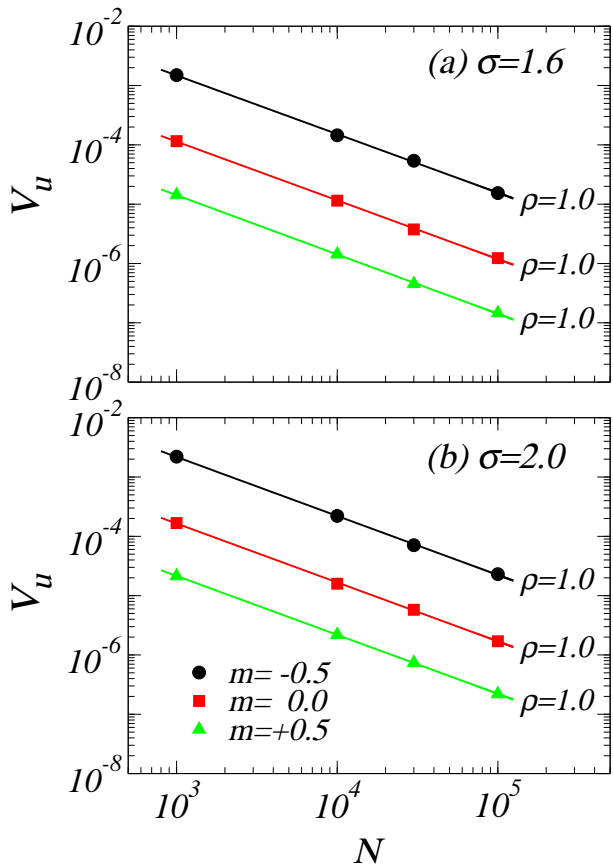}
\end{center}
\caption{\label{FIG5} (Color on line)   
  Variance $V_u(m)$ of the internal energy per spin $u(m)$ on random
  graphs with connectivity $z=4$ for selected values of $m$ as a
  function of system size. The number of disorder realizations is
  $10^4$. For the sake of clarity, the variances for $m=0$ and
  $m=0.5$ are divided by $10$ and $100$, respectively. The lines are
  fits to the form $V_u(m) \sim N^{-\rho}$, yielding $\rho \simeq 1.0$
  in all cases. }
\end{figure}

We now discuss the simulation results for the internal energy per spin
$u(m)$.  In Fig.~\ref{FIG5}, we plot on a log-log scale the variance
$V_u(m)= \langle u(m)^2 \rangle_M - \langle u(m)\rangle_M^2$ for
selected values of $m$ as a function of the system size $N$. Both
above and below $\sigma_c$, it is found that $V_u(m)$ decrease like
$1/N$, showing that the energy is a strongly self-averaging quantity,
a result that is not {\it a priori} obvious. Accordingly, we shall now
use the average value of $u(m)$ to compare with the exact results
obtained with the $H$-driven dynamics in the thermodynamic limit,
although $\langle u(m)\rangle_M$ may not be a well-defined physical
quantity, as remarked above. Rather, this must be considered as a
convenient way of suppressing sample-to-sample fluctuations.
\begin{figure}[htb]
\begin{center}
  \epsfig{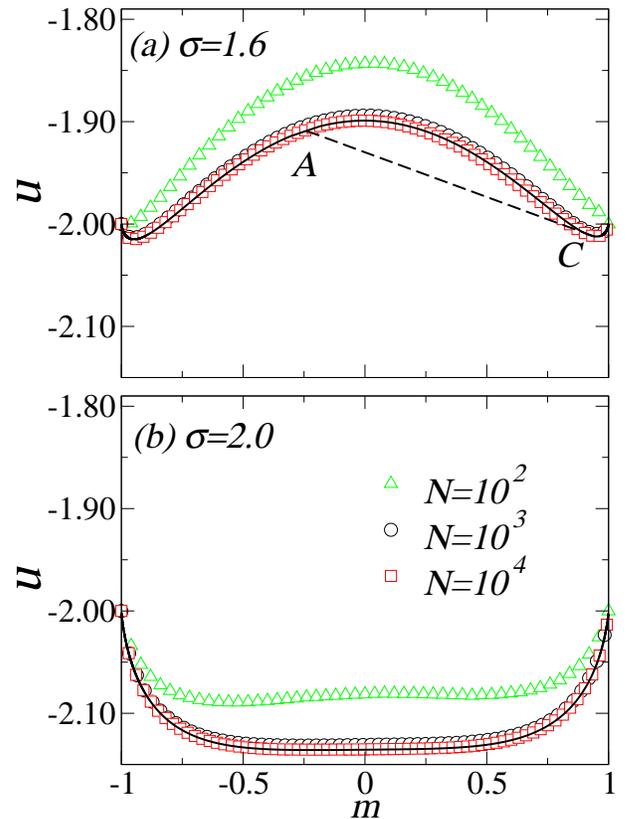}
\end{center}
\caption{\label{FIG6}  (Color on line)  
  Average internal energy per spin on the Bethe lattice with $z=4$
  below and above $\sigma_c$. The symbols are the results of the
  simulation of the $M$-driven algorithm on random graphs of different
  sizes with an average over $10^4$ disorder realizations. The solid
  line corresponds to the analytical expression given by Eq. (7).  The
  dashed line indicates the discontinuity associated to the infinite
  avalanche for $\sigma<\sigma_c$.}
\end{figure}

The comparison is performed in Fig.~\ref{FIG6} where $\langle
u(m)\rangle_H$ is obtained by plotting $\langle u\rangle_H$ as a
function of $\langle m\rangle_H$, the field $H$ being considered as a
parameter (see also Fig.~\ref{FIG1}). When $\sigma<\sigma_c$, there is
a jump in the magnetization and the corresponding discontinuity in
$\langle u(m)\rangle_H$ is represented by a dashed line, the solid
line representing the internal energy along the intermediate, unstable
part of the magnetization curve.  It can be seen hat the behavior of
$u(m)$ changes with $\sigma$. For large disorder, the energy has a
``double well'' structure whereas there is a single well when the
disorder is small. The change in the behavior occurs at $\sigma \simeq
2.0 $ and is therefore not related to the critical value of the
disorder at which the discontinuity in the $H$-driven magnetization
curve disappears.  Note moreover that the curves are not symmetric
with respect to $m=0$: there is indeed hysteresis when the
magnetization is increased from $-1$ or decreased from $+1$.
\begin{figure}[htb]
\begin{center}
  \epsfig{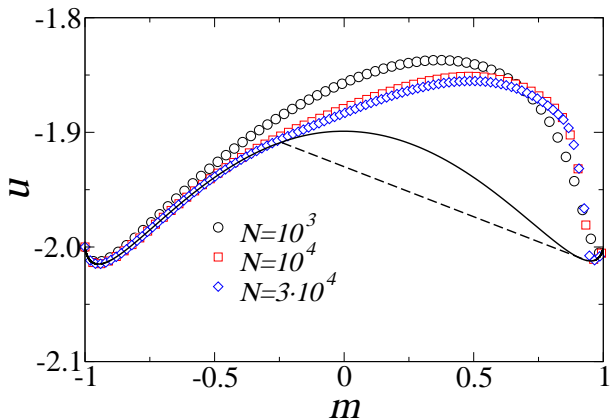}
\end{center}
\caption{\label{FIG7} (Color on line)
  Same as Fig.~\ref{FIG6}(a) when, inside an avalanche, one flips the spin that
  less decreases the energy. }
\end{figure}

The most remarkable feature in Fig.~\ref{FIG6} is that the average
internal energy obtained with the $M$-driven algorithm appears to
coincide with the analytical curve obtained from Eq. (\ref{ener}) in
the thermodynamic limit {\it even } when $m$ is in the range of the
macroscopic avalanche for $\sigma<\sigma_c$ (the agreement is better
than $10^{-3}$ for $N=10^4$).  Since this is a surprising result, we
have carefully checked the behavior as a function of the system size
(note incidentally that finite-size effets are not negligible in the
$H$-driven case as well: this is an issue that has not yet been
investigated, as far as we know).

When all avalanches are of microscopic size, the coincidence of the
energy along the two trajectories is due to the fact that the stable
states before and after the avalanche (and therefore all the unstable
states in-between) differ only by a finite (i.e., non-extensive)
number of spin-flips. Accordingly, the energy of these states cannot
differ by an extensive quantity and one has
\begin{equation}
\langle u(m)\rangle_M=\langle u(m)\rangle_M^{stable} = \langle u(m) \rangle_M^{unstable} 
\label{umh4}
\end{equation}
in the thermodynamic limit, as can be checked numerically. Moreover,
since both the energy and the magnetization are self-averaging
quantities, the averages at fixed $m$ or fixed $H$ yield the same
result. Therefore,
\begin{equation}
\langle u(m)\rangle_H\equiv \langle u(m)\rangle_H^{stable} =\langle u(m)\rangle_M^{stable} = \langle u(m) \rangle_M \ .
\label{umh5}
\end{equation}

It is more surprising that the equality $\langle u(m)\rangle_H=
\langle u(m) \rangle_M $ is also satisfied {\it inside} the
macroscopic avalanche if one uses the ``unphysical'' root of Eq.
(\ref{past}) to compute $\langle u(m)\rangle_H$ along the unstable
branch of the $H$-driven magnetization curve. We have no obvious
explanation for this result but we want to stress that it crucially
depends on the order in which the spins are flipped during an
avalanche. Indeed, as illustrated in Fig.~\ref{FIG7}, a different curve
$\langle u(m)\rangle_M$ is obtained if one decides for instance to
flip the spin that {\it less} decreases the energy. Therefore, the
$M$-driven dynamics that has been chosen is precisely the one that
yields agreement with the analytical solution computed in 
Ref.~\onlinecite{IOV2005}.  This suggests that behind the
probabilistic computation in Ref.~\onlinecite{DSS1997} there is perhaps some
hidden minimization principle that fixes unambiguously the trajectory
along the unstable branch.

\subsection{Statistical behavior of the output field}

\begin{figure}[htb]
\begin{center}
  \epsfig{file=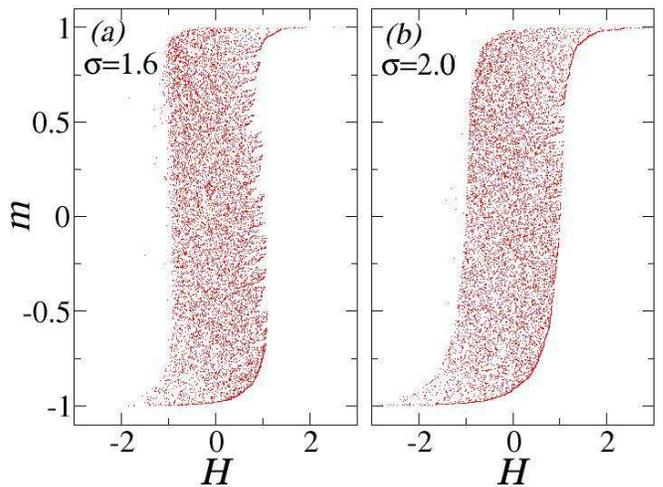,width=8.6cm,clip=}
\end{center}
\caption{\label{FIG8} (Color on line) Ascending branch of the $M$-driven trajectory in the field-magnetization plane. Data correspond to a single disorder realization on a random graph with connectivity $z=4$ ($N=10^4$).}
\end{figure}

As illustrated by Figs.~\ref{FIG2} and \ref{FIG8}, the output field
$H(m)$ defined by Eq. (5) displays a sporadic, discontinuous behavior
with magnetization. When a spin flips, the local field at the neigbors
is changed by $\pm 2$, and this is indeed the approximate size of the
fluctuations observed in Fig.~\ref{FIG8}. This of course does not
depend on the system size and the same behavior should be observed in
the thermodynamic limit. We shall come back to this important issue in
section \ref{Discussion}.  Another consequence of the definition of
the field as an extremal quantity is that it exhibits large
sample-to-sample fluctuations. In fact, it is found that the variance
does not decrease with $N$, which means that each sample behave
differently, even in the thermodynamic limit. It is however
instructive to study in detail the probability distribution $P_N(H;m)$
for different values of $m$ above and below $\sigma_c$.

\begin{figure}[htb]
\begin{center}
  \epsfig{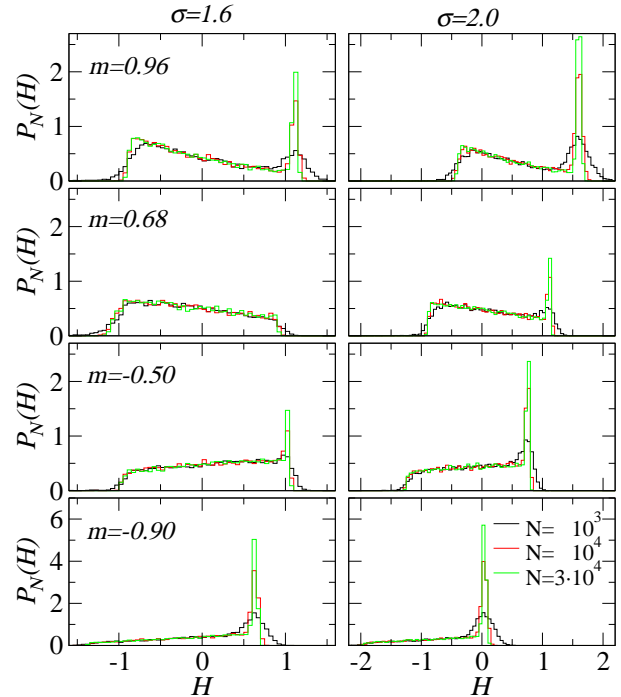}
\end{center}
\caption{\label{FIG9}  (Color on line)  Normalized histograms of the output field $H$ 
  for selected values of $m$ and different system sizes. The data
  correspond to $10^4$ disorder realizations. }
\end{figure}

The evolution of the normalized histograms as a function of system
size is shown in Fig.~\ref{FIG9}. One can see that the distributions
are wide and rather complicated. On the one hand, there is a
well-defined peak on the right-hand side of the histograms whose
height increases and width decreases as $N$ increases. This peak,
however, does not exist for $\sigma<\sigma_c$ when $m$ is in the range
of the infinite avalanche (for instance $m=0.68$ in the left panel of
the figure). On the other hand, there is another contribution which
extends over a finite range and which is almost size-independent: it
is responsible for the fact that the field is not self-averaging.  By
analyzing the sequence of microscopic states along each $M$-driven
trajectory, we have checked that these two contributions come from the
stable and unstable states, respectively. Since they are no other
stable states than those belonging to the $H$-driven magnetization
curve in the thermodynamic limit (as shown in Fig.~\ref{FIG3}), we therefore
conjecture that the distribution $P_N(H;m)$ has the following
asymptotic form:
\begin{equation}
 P_{\infty}(H;m) = Q(m) \delta (H-\hat{H}) + [1-Q(m)] w(H;m)
\label{eq2}
\end{equation}
where $\delta(H)$ is the Dirac function, $\hat{H}(m)$ is the field
along the magnetization curve (i.e. the field taken as a function of
the magnetization), and $w(H;m)$ is a continuous distribution on a
finite interval $\{H_{min}(m),H_{max}(m)\}$. Moreover, there is strong
numerical evidence that $H_{max}(m)=\hat{H}(m)$ for $\sigma>\sigma_c$
and for $\sigma<\sigma_c$ outside the range of the macroscopic
avalanche, whereas $H_{max}(m)<\hat{H}(m)$ inside the infinite
avalanche.

The statistical behavior of the field for a given value of the
magnetization is thus different above and below $\sigma_c$. For
$\sigma>\sigma_c$, the most probable value of $H(m)$ is the one
corresponding to the $H$-driven magnetization curve (the two protocols
thus give the same field-magnetization diagram), but there is a finite
probability that it takes a smaller value. For $\sigma<\sigma_c$ and
$m$ inside the range of the infinite avalanche, one has $Q(m)=0$ and
the delta peak disappears. In this case, the value of the field is
impredictable inside a finite interval.

\section{Results for the cubic lattice}
\label{3d}

Very similar results are obtained on the 3-D cubic lattice. In this
case, however, the $H$-driven behavior cannnot be treated exactly in
the thermodynamic limit and one must also perform simulations on
finite systems.

\begin{figure}[htb]
\begin{center}
  \epsfig{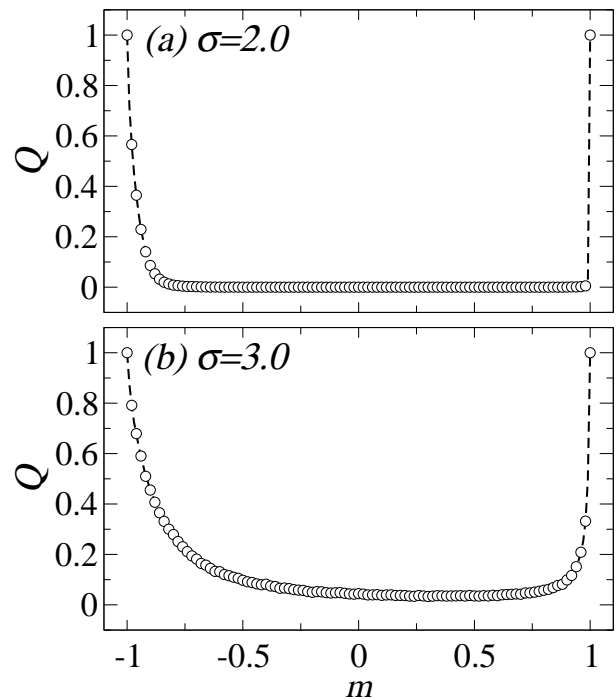}
\end{center}
\caption{\label{FIG10} Fraction $Q(m)$ of stable states along the $M$-driven 
trajectory on a cubic lattice (only the ascending branch is shown). The 
lattice size is $L=30$ and the average has been taken over $3\times 10^4$ 
disorder realizations. Lines are guides to the eye.}
\end{figure}

Fig.~\ref{FIG10} shows the fraction of stable states along the
$M$-driven trajectory. The curves are similar to the ones displayed in
Fig.~\ref{FIG4} except for a stronger asymmetry. Again, one finds that $Q(m)=0$
when $m$ is in the range of the $H$-driven macroscopic avalanche below
$\sigma_c$.

\begin{figure}[htb]
\begin{center}
  \epsfig{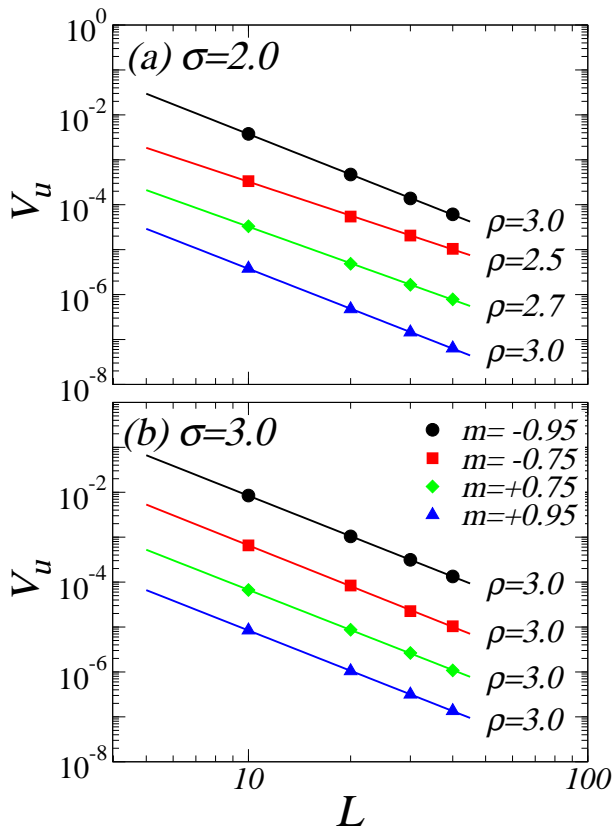}
\end{center}
\caption{\label{FIG11} (Color on line)
  Variance $V_u(m)$ of the internal energy per spin $u(m)$ on the
  cubic lattice for selected values of $m$ as a function of system
  size. Averages are performed over $3\times 10^4$ disorder
  realizations.  For the sake of clarity, the variances for $m=-0.75$,
  $m=0.75$ and $m=0.95$ are divided by $10$, $100$ and $1000$
  respectively.  The lines are fits to the form $V_u(m) \sim
  L^{-\rho}$. }
\end{figure}

Fig.~\ref{FIG11} shows that the internal energy obtained with the
$M$-driven algorithm is still a self-averaging quantity both above and
below $\sigma_c$ ($\sigma_c\simeq 2.2$). However, it seems that the
variance decreases slower than $1/L^3$ when $m$ is in the range of the
infinite avalanche, a behavior also observed Ref.\cite{IRV2006}
although the definition of the field is quite different.

\begin{figure}[htb]
\begin{center}
  \epsfig{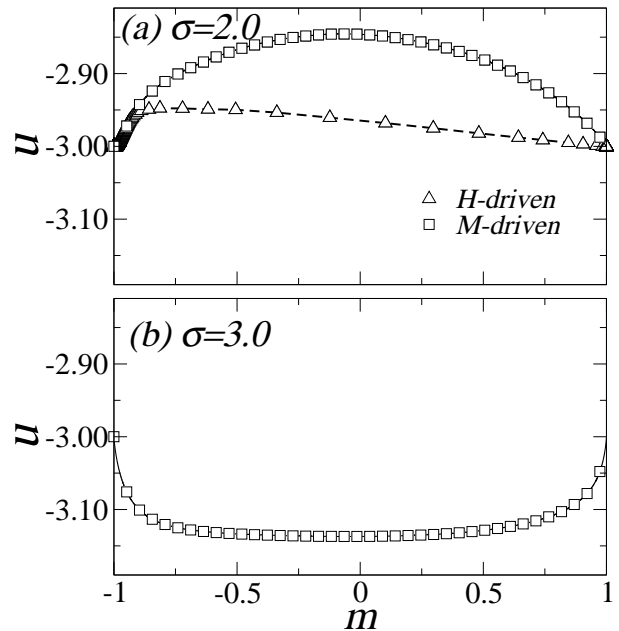}
\end{center}
\caption{\label{FIG12} 
  Average internal energy per spin on the cubic lattice below and
  above $\sigma_c$. The symbols represent the results of the
  simulation of a system with size $L=30$ with the $H$-driven and $M$-driven 
algorithms, as indicated. Averages are taken over $3 \times 10^4$ disorder
  realizations. Lines are guides to the eye. }
\end{figure}

\begin{figure}[htb]
\begin{center}
  \epsfig{file=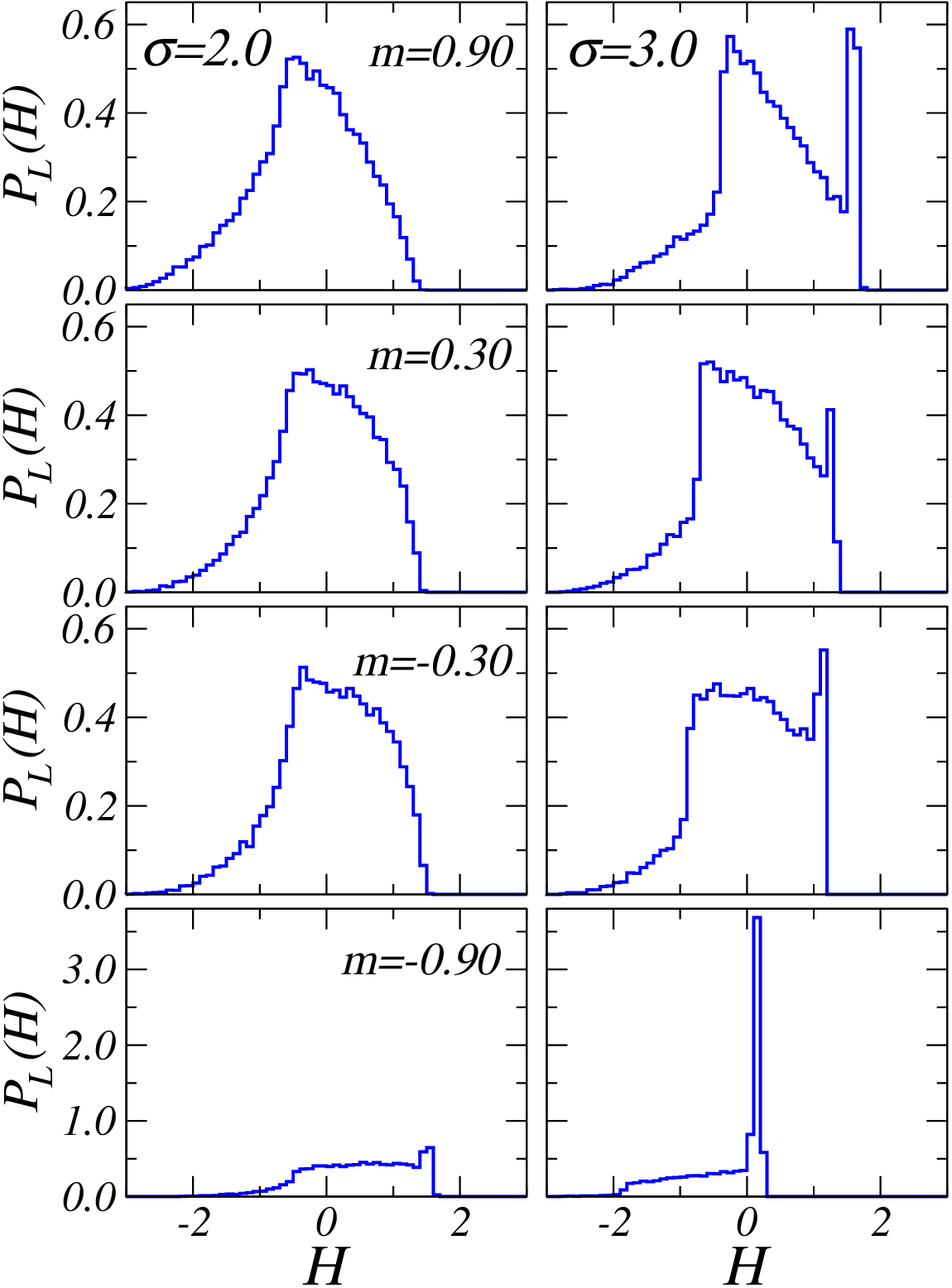,width=8cm,clip=}
\end{center}
\caption{\label{FIG13}  (Color on line) Normalized histograms of the output field $H$ 
  for selected values of $m$ and different system sizes. The data
  correspond to $3 \times 10^4$ disorder realizations of a system with size $L=30$.}
\end{figure}

The comparison between the two algorithms for the average internal
energy as a function of $m$ is performed in Fig.~\ref{FIG12}. Note
that there is again a double well structure at low disorder but the
two minima are very close to $m=\pm 1$ and hardly visible on the
figure.  There is only one minimum below $\sigma\approx 3$, a value
quite different from $\sigma_c$. It seems again that the two
algorithms give the same energy in the thermodynamic limit (outside
the infinite avalanche) but the finite-size effects are more important
than on the Bethe lattice.
Finally, the histograms of the field $H(m)$ are shown in Fig.~\ref{FIG13}. 
The overall behavior is similar the one displayed in Fig.~\ref{FIG9} 
and we thus conjecture that the asymptotic form of the probability
distribution is given by Eq. (18). Note however that the continuous
part $w(H;m)$ has a very different shape than on the Bethe lattice.

\section{Discussion}
\label{Discussion}

In the present paper, we have proposed a simple modification of the
standard single-spin-flip algorithm to study the magnetization-driven
RFIM at $T=0$. The dynamics consists in flipping the spins one by one,
choosing the spin with the largest local field. This allows to perform
a detailed comparison with the microscopic trajectory of the system in
the $H$-driven situation, above and below the critical disorder. It
turns out that the two trajectories share the same metastable states
in the thermodynamic limit, and we have computed the average fraction
of these states. An exact expression of this quantity has been
obtained in the case of the Bethe lattice. Numerical simulations show
that the two dynamics yield the same internal energy for a given value
of the magnetization outside the range of the macroscopic avalanche.
On the Bethe lattice, inside the macroscopic avalanche, the energy
obtained with the $M$-driven algorithm also coincides with the one
that can be computed analytically in the $H$-driven case, using the
solution of the self-consistent equations that describes the unstable
branch of the hysteresis loop.

The $M$-driven field-magnetization diagram exhibits some peculiar and
annoying features that are due to our definition of the output field
$H$ as $\Delta U/ \Delta M$: i) all closed loops have zero area,
implying that there is no dissipation in the system; ii) $H$ strongly
fluctuates with $m$ and these fluctuations are independent of the
system size; iii) the sample-to-sample fluctuations of $H$ also do not
decrease with the system size. We have shown that this problem is
related to the presence of a continuous part in the probability
distribution of $H$, which corresponds to the field associated to the
unstable states. We now discuss some possible modifications in the
definition of the dynamics or of the field.

The first one is to allow for an additional relaxation of the system
using the Kawasaki dynamics, which is the standard dynamics for a
situation with a conserved order parameter. Specifically, one could
imagine to first flip the spin with the largest local field (so to
change $M$) and then perform all possible exchanges between
nearest-neighbor spins of opposite sign that decrease the energy.
This procedure is certainly more in the spirit of the local mean-field
calculations that have been performed in Ref.\cite{IRV2006} at finite
temperature. Even without changing the definition of the field, one
may hope that the fluctuations of $H$ with $M$ will be weaker and will
perhaps decrease with $N$. However, preliminary simulations show that
the energy of some states cannot be decreased by exchanging
nearest-neighbor spins and that many of the final states are still
unstable with respect to the (Glauber) single-spin-flip dynamics. It
would be interesting to perform an extensive study in order to
understand if this behavior changes when increasing the system size.
On the other hand, it must be emphasize that the stable states are now
different from the ones visited by the $H$-driven dynamics. Moreover,
this procedure does not solve the problem of the definition of the
field (and the correponding absence of dissipation).

A second possibility is to keep the same dynamics as in this work, but
to change the definition of the output field.  Indeed, it is very
likely that the magnetization (or any other extensive variable) can
only be controlled at the macroscopic level within a certain
resolution. For instance, in an experiment performed at a constant
rate $dM/dt$, one probably measures not the instantaneous force (in
the present case $-f_i^{max}(m)$) but some average $\bar H$ over a
certain range $\Delta m$ (which could even depend on the driving
rate). In this case, one can easily check that all fluctuations are
suppressed in $\bar H$ in the thermodynamic limit since imposing a
fixed resolution $\Delta m$ implies to take averages over larger and
larger intervals $\Delta M$ when increasing $N$. $\bar H$ is then also
self-averaging.  One may also imagine that the apparatus that measures
the field (or the force) cannot ajusts itself to the force infinitely
fast or that there is some threshold value. Of course, in all these
cases, the results are machine-dependent. Carefull ``$M$-driven''
experiments with different set-ups and different driving rates are
thus needed in order to better resolve these issues.

Finally, from a theoretical point of view, one cannot discard the
possibility that there does not exist any satisfatory definition of
the output field when using Ising variables. An alternative approach
using continuous variables has been proposed in Ref.\cite{IRV2006}.
  
\acknowledgments The authors acknowledge fruitful discussions with F.
J. Perez-Reche, A. Planes, J. P. Sethna, Ll. Ma\~{n}osa and G. Tarjus.
This work has received financial support from CICyT (Spain), project
MAT2004-1291 and CIRIT (Catalonia), project 2005SGR00969.  Xavier Illa
acknowledges financial support from the Spanish Ministry of Education,
Culture and Sports. P. Shukla and M. L.  Rosinberg thank the
Generalitat de Catalunya for financial support (2004PIV2-2,
2005PIV1-17) and the hospitality of the ECM department of the
University of Barcelona.  The Laboratoire de Physique Th\'eorique de
la Mati\`ere Condens\'ee is the UMR 7600 of the CNRS.


\begin{thebibliography}{10}
  
\bibitem{SDP2004} J.P.Sethna, K.A.Dahmen, and O.Perkovi\'c in
  {\it The Science of Hysteresis}, edited by G.Bertotti and I.Mayergoyz, Elsevier (2004).
  
\bibitem{MB2005} Q.Michard and J.P.Bouchaud, Eur. Phys. J. B {\bf
    47}, 151 (2005).
  
\bibitem{OW1998} K.Otsuka and C.M.Wayman in {\it Shape Memory
    Materials}, edited by K.Otsuka and C.M.Wayman, Cambridge
  University Press, Cambridge (1998).
  
\bibitem{G1977} W.Grosse-Nobis, J. Mag. Mag. Mater.  {\bf 4}, 247
  (1977).

\bibitem{BMPRV2006} E.Bonnot, Ll.Ma\~nosa, A.Planes, R.Romero, and
  E.Vives (in preparation).
  
\bibitem{SDKKRS1993} J.P.Sethna, K.Dahmen, S.Kartha, J.A.Krumhansl, 
B.W.Roberts, and J.D.Shore, Phys. Rev. Lett. {\bf
    70}, 3347 (1993).
  
\bibitem{IRV2006} X.Illa, M.L.Rosinberg, and E.Vives, preprint
  cond-mat/0607069.
  
\bibitem{DSS1997} D.Dhar, P.Shukla, and J.P.Sethna, J. Phys. A:
  Math. Gen. {\bf 30} 5239 (1997).
  
\bibitem{SSD2000} S.Sabhapandit, P.Shukla, and D.Dhar, J. Stat.
  Phys.  {\bf 98} 103 (2000).
  
\bibitem{IOV2005} X.Illa, J.Ort\'{\i}n and E.Vives, Phys. Rev. B {\bf
    71}, 184435 (2005).
  
\bibitem{Perkovic1999} O.Perkovi{\'c}, K.A.Dahmen and J.P.Sethna,
  Phys. Rev. B {\bf 59}, 6106 (1999).
  
\bibitem{PV2003} F.J.Perez-Reche and E.Vives, Phys. Rev. B {\bf
    67}, 134421 (2003).
  
\bibitem{Vives2005} E.Vives, M.L.Rosinberg, and G.Tarjus, Phys.
  Rev. B {\bf 71}, 134424 (2005).
  
\bibitem{DRT2005} F.Detcheverry, M.L.Rosinberg, and G.Tarjus, Eur.
  Phys. J. B {\bf 44}, 327 (2005).
  
\bibitem{S1992} K.Sneppen, Phys. Rev. lett. {\bf 69}, 3539 (1992); 
  P.Bak and K.Sneppen, Phys. Rev. lett. {\bf 71}, 4083 (1992), 
  M.Paczuski, S.Maaslov, and P.Bak, Phys. Rev. E {\bf 53}, 414 (1996).
  

  
\bibitem{B1959} R.Brout, Phys. Rev. {\bf 115}, 824 (1959).
  
\bibitem{BH1988} K.Binder and D.W.Heermann, {\it Monte Carlo
    Simulations in Statistical Physics}, Springer-Verlag, Berlin
  (1988).
  
\bibitem{ISV2006} X.Illa, P.Shukla, and E.Vives, Phys. Rev. B {\bf
    73}, 092414 (2006).

\end{thebibliography}

\end{document}